# Comparative Study Between Dispersive and Non-Dispersive Dielectric Permittivity in Spectral Remittances of Chiral Sculptured Zirconia Thin Films

Ferydon Babaei[1,*] and Hadi Savaloni[2]

1) Department of Physics, University of Qom, Qom, Iran.
2) Department of Physics, University of Tehran, North-Kargar Street, Tehran, Iran.
*) Corresponding author: Tel: +98 251 2853311; Fax: +98 251 2854972; Email: fbabaei@qom.ac.ir

## Abstract

The transmission and reflection spectra from a right-handed chiral sculptured zirconia thin film are calculated using the piecewise homogeneity approximation method and the Bruggeman homogenization formalism by considering that the propagation of both dispersive and non-dispersive dielectric function occurs for axial and non-axial states. The comparison of spectral results shows that the dispersion of the dielectric function has a considerable effect on the results. In axial excitation of cross-polarized reflectances and co-polarized transmittances the dispersion effect becomes more pronounced at wavelengths further away from the homogenization wavelength. This is also true in case of non-axial excitation of circular transmittances, while there are considerable differences for cross-polarized reflectances where (wavelength) the first Bragg peak occurs. At wavelengths in the vicinity of the homogenization wavelength the dispersion effect of the dielectric function in $R_{RR}$ becomes more significant.

## 1. Introduction

In order to obtain the transmission and reflection spectra for sculptured thin films, usually the relative dielectric constant of the film material at a certain frequency is considered, then the relative permittivity scalars are estimated, using the Bruggeman homogenization formalism. These scalar quantities are assumed constant in the procedure of obtaining the transmission and reflection spectra at all frequencies [1-3]. However, we know that the relative dielectric constant varies with the frequency. In addition, in the main body of the literature on the dielectric dispersion function effect on remittances (reflection and transmission) of sculptured thin films, the simple single-resonance Lorentzian model is used [4-6], while in order to be able to define the oscillator strengths, resonance wavelengths, and absorption linewidths, one should have a good knowledge of the oscillatory and quantum behavior of the thin film. In addition there is an argument that all thin films may not obey from the single-resonance Lorentzian model and are composed of a double-resonance [7] or multiple-resonance [8,9] systems. In order to avoid such complications we have implemented the experimental results of the refractive index of zirconia thin film at each wavelength to calculate the dispersion of the dielectric constant. **It should also be mentioned here that the refractive index for zirconia in the wavelength region examined in this work has only real values and the imaginary part is zero [10].**
In our earlier works, we have reported the reflectance and transmittance from an axially [11] and a non-axially [12] excited chiral sculptured zirconia thin film. In this paper we report on the influence of dispersive and non-dispersive dielectric functions on the circular reflectance and transmittance spectra in both axial and non-axial propagation states by using the experimental data of refractive index of zirconia thin film.

## 2. Theory

Consider that a region ($0 \leq z \leq d$) in space is occupied by a chiral sculptured thin film (CSTF) and that this film is being excited by a plane wave which propagates with an angle $\theta_{inc}$ relative to z axis and angle $\psi_{inc}$ relative to x axis in xy plane. The phasors of incident, reflected and transmitted electric fields are given as [13]:

$$\begin{cases} \underline{E}_{inc}(\underline{r}) = [(\frac{i\underline{s}-\underline{p}_+}{\sqrt{2}})a_L - (\frac{i\underline{s}+\underline{p}_+}{\sqrt{2}})a_R]e^{i\underline{K}_0 \cdot \underline{r}}, z \leq 0 \\ \underline{E}_{ref}(\underline{r}) = [-(\frac{i\underline{s}-\underline{p}_-}{\sqrt{2}})r_L + (\frac{i\underline{s}+\underline{p}_-}{\sqrt{2}})r_R]e^{-i\underline{K}_0 \cdot \underline{r}}, z \leq 0 \\ \underline{E}_{tr}(\underline{r}) = [(\frac{i\underline{s}-\underline{p}_+}{\sqrt{2}})t_L - (\frac{i\underline{s}+\underline{p}_+}{\sqrt{2}})t_R]e^{i\underline{K}_0 \cdot (\underline{r}-d\underline{u}_z)}, z \geq d \end{cases} \quad (1)$$

The magnetic field's phasor in any region is given as:

$$\underline{H}(\underline{r}) = (i\omega\mu_0)^{-1}\underline{\nabla} \times \underline{E}(\underline{r})$$

where $(a_L, a_R)$, $(r_L, r_R)$ and $(t_L, t_R)$ are the amplitudes of incident plane wave, and **reflected and transmitted** waves with left- or right-handed polarizations. We also have;

$$\begin{cases} \underline{r} = x\underline{u}_x + y\underline{u}_y + z\underline{u}_z \\ \underline{K}_0 = K_0(\sin\theta_{inc}\cos\psi_{inc}\underline{u}_x + \sin\theta_{inc}\sin\psi_{inc}\underline{u}_y + \cos\theta_{inc}\underline{u}_z) \end{cases} \quad (2)$$

where $K_0 = \omega\sqrt{\mu_0\varepsilon_0} = 2\pi/\lambda_0$ is the free space wave number, $\lambda_0$ is the free space wavelength and $\varepsilon_0 = 8.854 \times 10^{-12} Fm^{-1}$ and $\mu_0 = 4\pi \times 10^{-7} Hm^{-1}$ are the permittivity and permeability of free space (vacuum), respectively. The unit vectors for linear polarization parallel and normal to the incident plane, $\underline{s}$ and $\underline{p}$, respectively are defined as:

$$\begin{cases} \underline{s} = -\sin\psi_{inc}\underline{u}_x + \cos\psi_{inc}\underline{u}_y \\ \underline{p}_\pm = \mp(\cos\theta_{inc}\cos\psi_{inc}\underline{u}_x + \cos\theta_{inc}\sin\psi_{inc}\underline{u}_y) + \sin\theta_{inc}\underline{u}_z \end{cases} \quad (3)$$

and $\underline{u}_{x,y,z}$ are the unit vectors in Cartesian coordinates system.

The reflectance and transmittance amplitudes can be obtained, using the continuity of the tangential components of electrical and magnetic fields at two interfaces, $z = 0$ and $z = d$, and solving the algebraic matrix equation [13]:

$$\begin{bmatrix} i(t_L - t_R) \\ -(t_L + t_R) \\ 0 \\ 0 \end{bmatrix} = \left[\underline{\underline{K}}(\theta_{inc}, \psi_{inc})\right]^{-1} \cdot \left[\underline{\underline{B}}(d,\Omega)\right] \cdot \left[\underline{\underline{M}}'(d,\Omega,\kappa,\psi_{inc})\right] \cdot \left[\underline{\underline{K}}(\theta_{inc}, \psi_{inc})\right] \cdot \begin{bmatrix} i(a_L - a_R) \\ -(a_L + a_R) \\ -i(r_L - r_R) \\ (r_L + r_R) \end{bmatrix} \quad (4)$$

different terms and parameters of this equation are given in detail by Venugopal and Lakhtakia (see equations (2-25), (2-26) in reference [13]).

In order to obtain $[\underline{\underline{M}}'(d,\Omega,\kappa,\psi_{inc})]$ the piecewise homogeneity approximation method [14] is used. In this method the CSTF is divided into N (a big enough number) very thin layers with a thickness of $h = d/N$ (5 nm will suffice).

Once the transmittance and the reflectance amplitudes are obtained from Eq. (4), then we can obtain the reflectance and transmittance coefficients as:

$$\begin{cases} r_{i,j} = \dfrac{r_i}{a_j} \\ t_{i,j} = \dfrac{t_i}{a_j} \end{cases}, i, j = L, R \quad (5)$$

The transmittance and reflectance are obtained from:

$$\begin{cases} R_{i,j} = |r_{i,j}|^2 \\ T_{i,j} = |t_{i,j}|^2 \end{cases}, i, j = L, R \quad (6)$$

### 3. Numerical results and discussion

We consider that a right-handed zirconia sculptured thin film with a thickness $d$ in its bulk state has occupied the free space. The relative permittivity scalars $\varepsilon_{a,b,c}$ in this sculptured thin film were obtained using the Bruggeman homogenization formalism [15,16]. In this formalism, the film is considered as a two phase composite, vacuum phase and the inclusion phase. These quantities are dependent on different parameters, namely, columnar form factor, fraction of vacuum phase (void fraction), the wavelength of free space and the refractive index $n_s(\omega)+ik(\omega)$ of the film's material (inclusion). Each column in the STF was assumed to consist of a string of small and identical ellipsoids and are electrically small (i.e., small in a sense that their electrical interaction can be ignored). Therefore [11,12]:

$$\varepsilon_\sigma = \varepsilon_\sigma(\varepsilon_s(\lambda_\circ), f_v, \gamma_\tau^s, \gamma_b^s, \gamma_\tau^v, \gamma_b^v), \qquad \sigma = a,b,c \qquad (7)$$

where $f_v$ is the fraction of void phase, $\varepsilon_s(\omega)=(n_s(\omega)+ik(\omega))^2$ is the relative dielectric permittivity, $\gamma_\tau^{s,v}$ is one half of the long axis of the inclusion and void ellipsoids, and $\gamma_b^{s,v}$ is one half of the small axis of the inclusion and void ellipsoids. In all calculations the following parameters were fixed; $\gamma_b^s = 2$, $\gamma_\tau^s = 20$, $\gamma_b^v = 1$ $\gamma_\tau^v =$, $f_v = 0.6$, $\chi = 30°$, $\Omega = 162 nm$, $d = 40\Omega$, and a range of wavelengths $\lambda_0 \in (250nm - 850nm)$ was considered, where the real refractive index **of zirconia in its bulk state (Fig. 1)** varies from 2.64599 to 2.17282 for the lowest wavelength to highest wavelength, respectively [10]. The main parameters chosen in this work, namely $\gamma_\tau^s$, $f_v$, $\Omega$, $d$ are very similar to those reported by Sherwin et al. [16] for titanium oxide. The difference between our other parameters and those of Sherwin et al [16] may be explained on the basis that in our work we have not reported any experimental results for zirconia, while Sherwin et al obtained experimental data and fitted that to their theoretical work. This can be admittedly considered as one of the weaknesses of our work, which is being considered for future studies. **It should be noted that the imaginary part of the refractive index for zirconia in this range of the wavelengths is zero (Fig.1), hence dissipation can be ignored. In addition it is worthwhile to clarify that the bulk data for zirconia presented in reference [10] is** **incremented in 10 nms. In order to carry out our calculations for each of these wavelengths it was necessary to iterate the Bruggeman equation 12 times.**

In each plot of Figs. 2, 4 and 6 four spectra (curves) are depicted. In curve (i) the dispersion of dielectric function is included in the Bruggman homogenization formalism (i.e., homogenization is implemented for each wavelength). In curves (ii), (iii) and (iv), the homogenization is performed for $\lambda_0 < \lambda_{0,dis}^{Br}$, $\lambda_0 = \lambda_{0,dis}^{Br}$, and $\lambda_0 > \lambda_{0,dis}^{Br}$ ($\lambda_{0,dis}^{Br}$ is the Bragg wavelength when dispersion of dielectric function is taken into account), respectively. The Bragg wavelength in Figs.2, 4 and 6 is 480 nm, 420 nm and 410 nm, respectively, which was obtained using Bruggeman formalism including dispersion function and the data is presented as curve (i) in each figure. The lower and higher wavelengths in each case are given in the figure captions. Therefore, in the latter, the permittivity scalars remain the same for other wavelengths. In each plot of Figs.3, 5, and 7, three spectra are depicted (curves (i-ii), (i-iii) and (i-iv)). These spectra show the difference between the values obtained in Figs.2, 4 and 6 for dispersed and non-dispersed states of reflectance and transmittance. These differences are calculated using:

$$\Delta R_{p,q} = \left| R_{p,q}(\lambda_0^i) - R_{p,q}(\lambda_0^j) \right|, \quad p,q = L, R\; ; i = curve(i), j = curves(ii,,iii.iv) \quad (8)$$

where $R_{p,q}(\lambda_0^i)$ and $R_{p,q}(\lambda_0^j)$ present the homogenization over the whole wavelength region and homogenization at a certain wavelength, respectively.

In Fig. 2 the circular reflectance and transmittance spectra, and in Fig. 3 the differences of the values obtained in Fig. 2 for circular reflectance and transmittance between dispersive and non-dispersive states for a right-handed zirconia CSTF as a function of wavelength $\lambda_0$ for $\theta_{inc} = \psi_{inc} = 0$ in axial propagation state (z axis) are given.

It is well known that for an axially excited CSTF, $R_{RL} = R_{LR}$ and $T_{RL} = T_{LR}$; i. e., the corss-polarized reflected and transmitted intensities do not show any dependence on the circular polarization state of the incident plane wave. This is a consequence of the relative permittivity dyadic [1,2] of the CSTF being symmetric [17].

The results presented in Figs.2 and 3 can be interpreted as follows:

*a) reflectance:*

1) in the $R_{LL}$ plot, since the structural **handedness** of the thin film is not the same as the polarization of the incidence plane wave, Bragg peaks are smaller and there is no considerable difference between dispersive (curve (i)) and non-dispersive states (curves (ii), (iii) and (iv)). It can be observed that $\Delta R_{LL}$ is negligible.

2) in the $R_{RL}$ plot, it can be seen that there exist a relatively a considerable difference between dispersive and non-dispersive states $\Delta R_{RL}$, and this difference increases by moving further away from the homogenization wavelength (i.e., the wavelength at which the homogenization is being performed). The homogenization wavelength for curves (ii), (iii) and (iv) were 430 nm, 480 nm, and 530 nm, respectively.

3) in the $R_{RR}$ plot, owing to the same **structural handedness in** the structural direction in the thin film and the polarization of the incidence plane wave, circular Bragg peaks occur vividly in the Bragg region. The difference $\Delta R_{RR}$ between the spectra at wavelengths near to homogenization wavelengths is small and becomes zero at homogenization wavelength. In order to clearly observe how the Bragg regimes are affected through theses processes, the $R_{RR}$ and $\Delta R_{RR}$ plots in the wavelength region of 450 to 550 nm are given in Fig. 4(a-b). In Fig. 4(b), it can be observed that the Bragg regime is most affected at lower wavelengths than that of Bragg.

b) *transmittance:*

1) a right-handed chiral sculptured thin film transmits the **LCP** plane wave almost completely (Fig.2: $T_{LL}$ plot). It can be observed that there exists a considerable difference between dispersed and non-dispersed states. This difference is more pronounced at wavelengths further away from the homogenization wavelength (Fig.3: $\Delta T_{LL}$ plot).

2) in $T_{RL}$ plot, the transmittance spectrum in the Bragg region stands at higher values than those outside Bragg region. No difference can be observed between dispersed and non-dispersed spectra (Fig.3: $\Delta T_{RL}$ plot).

3) **in $T_{RR}$ plot, in the Bragg region, the transmitted spectrum is minimized.** At wavelengths further away from the homogenization wavelength, the difference between the dispersed and the non-dispersed states become more pronounced (Fig.3: $\Delta T_{RR}$ plot).

In Fig. 5 the circular remittances and in Fig. 6 the differences of the values obtained in Fig. 5 for circular reflectances and transmittances between the dispersive and the non-dispersive states for a right-handed zirconia CSTF as a function of wavelength $\lambda_0$ for $\theta_{inc} = 45°$, $\psi_{inc} = 0°$ in the non-axial propagation state (xy plane) are given.

The results given in Fig. 5 and Fig. 6 may be described as follows:

a) *reflectance*

1) in $R_{LL}$ plot the circular Bragg phenomenon character is not obvious, but there is a considerable difference between dispersed and non-dispersed states, and at wavelengths further away from the homogenization wavelength this difference becomes more pronounced (Fig. 6: $\Delta R_{LL}$ plot). The homogenization wavelength for curves (ii), (iii) and (iv) were chosen as 370 nm, 420nm and 470 nm, respectively.

2) In $R_{RL}$ plot, the first and second Bragg peaks are clearly observed, while the higher order peaks occur at shorter wavelengths with much less strength, hence they cannot clearly be observed. There is a considerable difference between dispersed and non-dispersed states, which is more pronounced at longer wavelengths (i.e., where the first Bragg peak appears) (Fig. 6: $\Delta R_{RL}$ plot).

3) The discussion and interpretation of the results given for $R_{RL}$ (Fig. 5) and $\Delta R_{RL}$ (Fig. 6) are applicable to $R_{LR}$ (Fig. 5) and $\Delta R_{LR}$ (Fig. 6) with a little difference between spectral values. This shows that the CSTF discriminates between right circularly and left circularly polarized plane waves.

In $R_{RR}$ plot the occurrence of circular Bragg phenomenon is more distinguishable. Also the difference between dispersed and non-dispersed states (curves (ii), (iii) and (iv)) in the Bragg region is more pronounced (Fig. 6: $\Delta R_{RR}$ plot). The reason for small difference between the non-dispersed state (curve (iii)) and the dispersed state (curve (i)) is due to the fact that in curve (iii) the homogenization has been performed at only 420 nm wavelength, which is the same wavelength for which the reflectance spectrum in dispersed state is maximum. In order to observe The influence of the incident angle, $\theta_{inc}$, on the Bragg regimes, the $R_{RR}$ and $\Delta R_{RR}$ plots in the wavelength region of 400 to 500 nm are given in Fig. 7(a-b). It can be observed in Fig. 7(b) that similar to Fig. 4(b), the Bragg regime is most affected at lower wavelengths than that of Bragg. However, comparison of Fig. 7(b) with Fig. 4(b) shows that the influence of the change in the incident angle from zero degree to 45 degrees has affected the Bragg regime by a factor of ten.

*b) transmittance:*

1) $T_{LL}$ spectra are disordered and the effect of dispersion of dielectric function can be clearly distinguished in the regions far away from the homogenization wavelength (Fig. 6: $\Delta T_{LL}$ plot).

2) There exist a trough in the $T_{RL}$ spectra which corresponds to the second Bragg peak. The influence of the dispersion of dielectric function at longer wavelengths is more pronounced (Fig. 6: $\Delta T_{RL}$ plot).

3) The discussion and interpretation of the results given above for $T_{RL}$ and $\Delta T_{RL}$ are applicable to $T_{LR}$ and $\Delta T_{LR}$ with a little difference between spectral values. This shows that the CSTF discriminates between right circularly and left circularly polarized plane waves.

4) Two troughs can be observed in $T_{RR}$ spectra. The difference between the dispersed and non-dispersed dielectric functions become more vivid at wavelengths further away from the homogenization wavelength (Fig. 6: $\Delta T_{RR}$ plot).

In Fig. 8 the circular remittances and in Fig. 9 the differences of the values obtained in Fig. 8 for circular reflectance and transmittance between dispersive and non-dispersive states for a right-handed zirconia CSTF as a function of wavelength $\lambda_0$ for $\theta_{inc} = 45°$, $\psi_{inc} = 90°$ in non-axial propagation state (yz plane) are given.

The comparison of Figs. 8 and 9 with Figs. 5 and 6, respectively shows that in fact the obtained spectra are almost similar and the only difference is that in Figs. 8 and 9 the homogenization for curves (ii), (iii) and (iv) is performed at 360 nm, 410 nm and 460 nm wavelengths. The fundamental difference between dispersive and non-dispersive states is in $\Delta R_{RR}$ (Fig. 9 curve (i-ii)) which is clearly distinguishable at homogenization wavelength.

Fig. 10 shows the plots of $R_{RR}$ and $\Delta R_{RR}$ in the wavelength region of 400 to 500 nm. Fig. 10(b) again shows the influence of the shorter wavelengths on the Bragg regime, while higher wavelengths have lesser effect,

In summary, in this work by using the Bruggeman homogenization formalism and the piecewise homogeneity approximation method we have been able to show the influence of the dispersion of dielectric function in reflectance and transmittance spectra of circularly polarized plane waves from a right-handed zirconia CSTF for both axial and non-axial propagation states. The influence of dispersion effect on axial excitation of cross-polarized reflectances and co-polarized transmittances becomes more detectable at wavelengths further away from the homogenization wavelength. For non-axial excitation of circular transmittances similar results to those of axial excitation are obtained. There exist fundamental differences for cross-polarized reflectances where (wavelength) the first Bragg peak occurs. The dispersion effect of the dielectric function in $R_{RR}$ becomes more significant at wavelengths near the homogenization wavelength.

## 4. Conclusions

The influence of the dispersion of dielectric function in the remittances spectra of circularly polarized plane waves from a right-handed zirconia CSTF is reported, using the piecewise homogeneity approximation method and the Bruggeman homogenization formalism at each given frequency for both axial and non-axial propagation states. This was carried out by considering the refractive index of zirconia at each given frequency, individually, in the frequency range of 250 to 850 nm in the homogenization formalism. Therefore in this way dispersion of the dielectric function was introduced into our calculations. This method directly takes advantage from the experimental relative dielectric constant of thin film and avoids the use of simple dispersion model known as single-resonance Lorentzian model, because it is believed that not all thin film systems may obey the single-resonance Lorentzian model, but may be composed of a double-resonance or multiple-resonance systems.

## Acknowledgements

This work was carried out with the support of the University of Tehran and the Iran National Science Foundation (INSF).

**Figure captions**

Figure 1. The refractive index of pure bulk zirconia, showing that the imaginary part is zero in the wavelength region shown.

Figure 2. Reflectance and transmittance spectra from a right-handed zirconia CSTF as a function of wavelength $\lambda_0$ for $\theta_{inc} = \psi_{inc} = 0°$, in axial propagation (z axis). a) reflectance; b) transmittance. The calculations performed for curves (i) to (iv) according to the following conditions:
i) $n_s(250nm, 850nm) \in [2.64599, 2.17482]$, ii) $n_s(430nm) = 2.24993$, iii) $n_s(480nm) = 2.22853$, iv) $n_s(530nm) = 2.22117$.

Figure 3. The differences obtained between dispersive and non-dispersive states of the results presented in Fig. 1 for reflectances and transmittances.

Figure 4. Plots of a) $R_{RR}$ and b) $\Delta R_{RR}$ as in figures 2 and 3, but drawn in the wavelength range of 450 to 550 nm.

Figure 5. Reflectance and transmittance spectra from a right-handed zirconia CSTF as a function of wavelength $\lambda_0$ for $\theta_{inc} = 45°, \psi_{inc} = 0°$, in non-axial propagation (xz plane). a) reflectance; b) transmittance. The calculations performed for curves (i) to (iv) according to the following conditions:
i) $n_s(250nm, 850nm) \in [2.64599, 2.17482]$, ii) $n_s(370nm) = 2.30807$, iii) $n_s(420nm) = 2.25695$, iv) $n_s(470nm) = 2.23125$.

Figure 6. The differences obtained between dispersive and non-dispersive states of the results presented in Fig. 3 for reflectances and transmittances.

Figure 7. Plots of a) $R_{RR}$ and b) $\Delta R_{RR}$ as in figures 5 and 6, but drawn in the wavelength range of 400 to 500 nm.

Figure 8. Reflectance and transmittance spectra from a right-handed zirconia CSTF as a function of wavelength $\lambda_0$ for $\theta_{inc} = 45°, \psi_{inc} = 90°$, in non-axial propagation (yz plane). a) reflectance; b) transmittance. The calculations performed for curves (i) to (iv) according to the following conditions:
i) $n_s(250nm, 850nm) \in [2.64599, 2.17482]$, ii) $n_s(360nm) = 2.32186$, iii) $n_s(410nm) = 2.26501$, iv) $n_s(460nm) = 2.23465$.

Figure 9. The differences obtained between dispersive and non-dispersive states of the results presented in Fig. 5 for reflectances and transmittances

Figure 10. Plots of a) $R_{RR}$ and b) $\Delta R_{RR}$ as in figures 8 and 9, but drawn in the wavelength range of 400 to 500 nm.

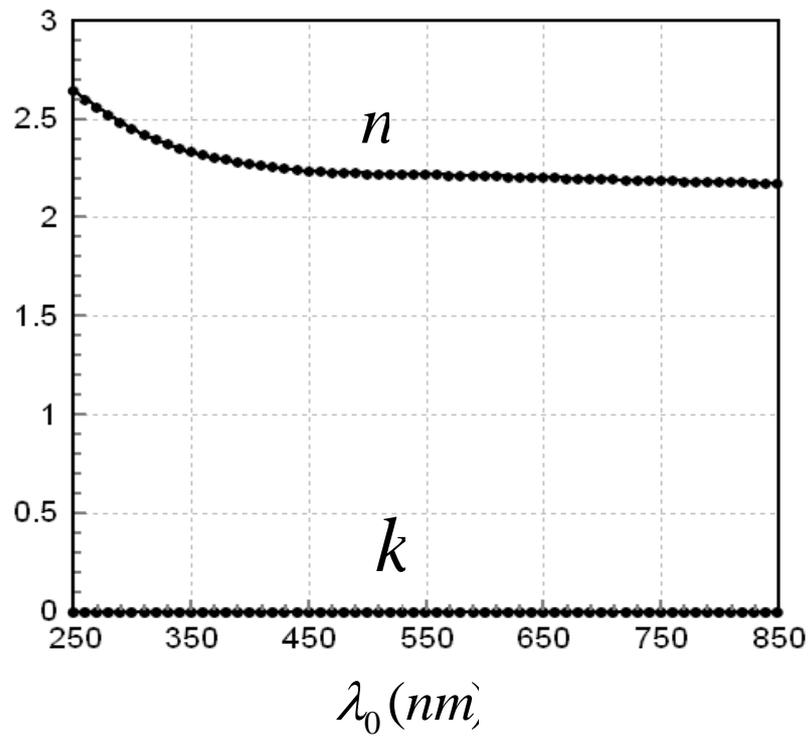

**Fig.1; F. Babaei and H. Savaloni**

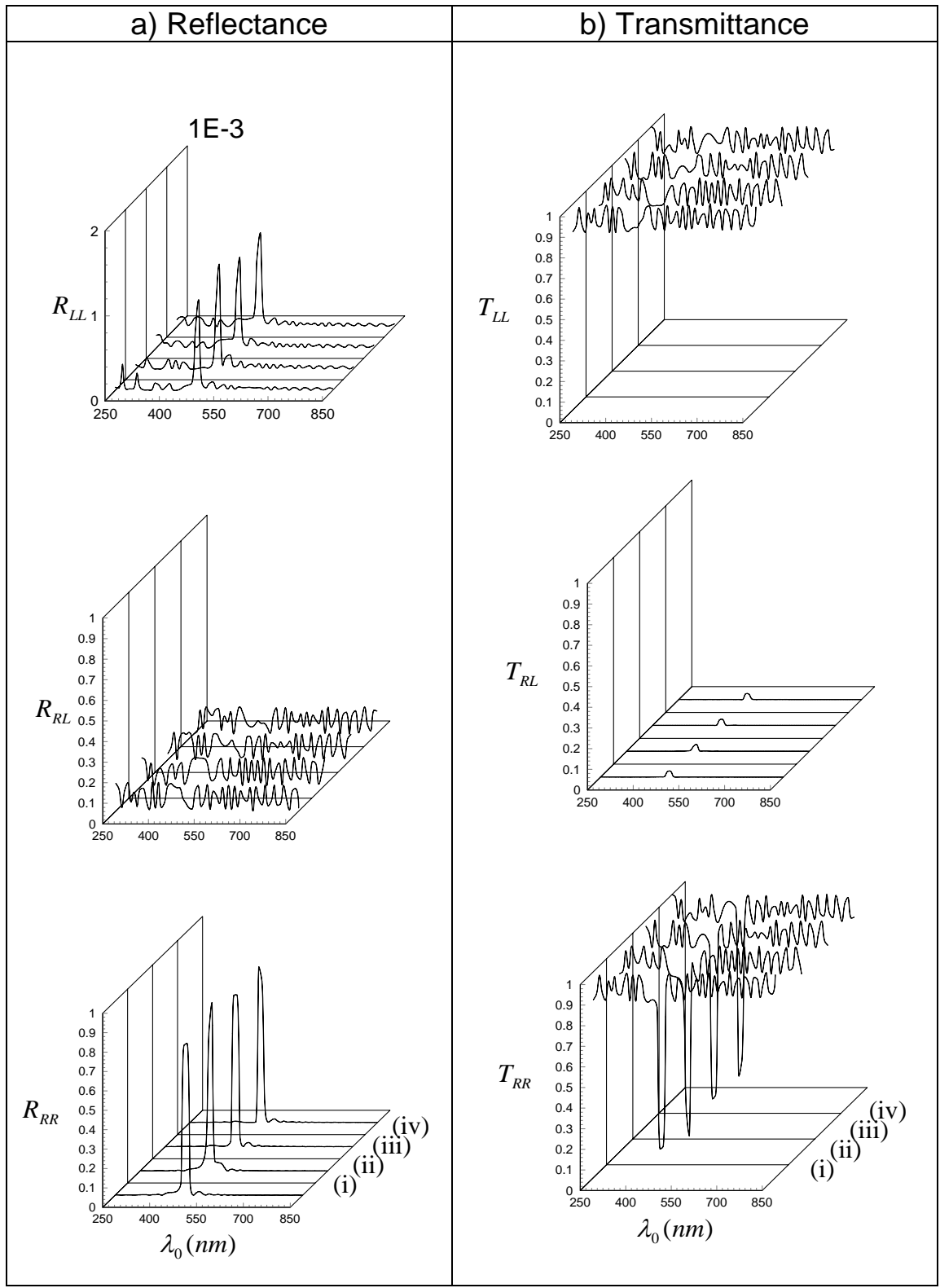

**Fig.2; F. Babaei and H. Savaloni**

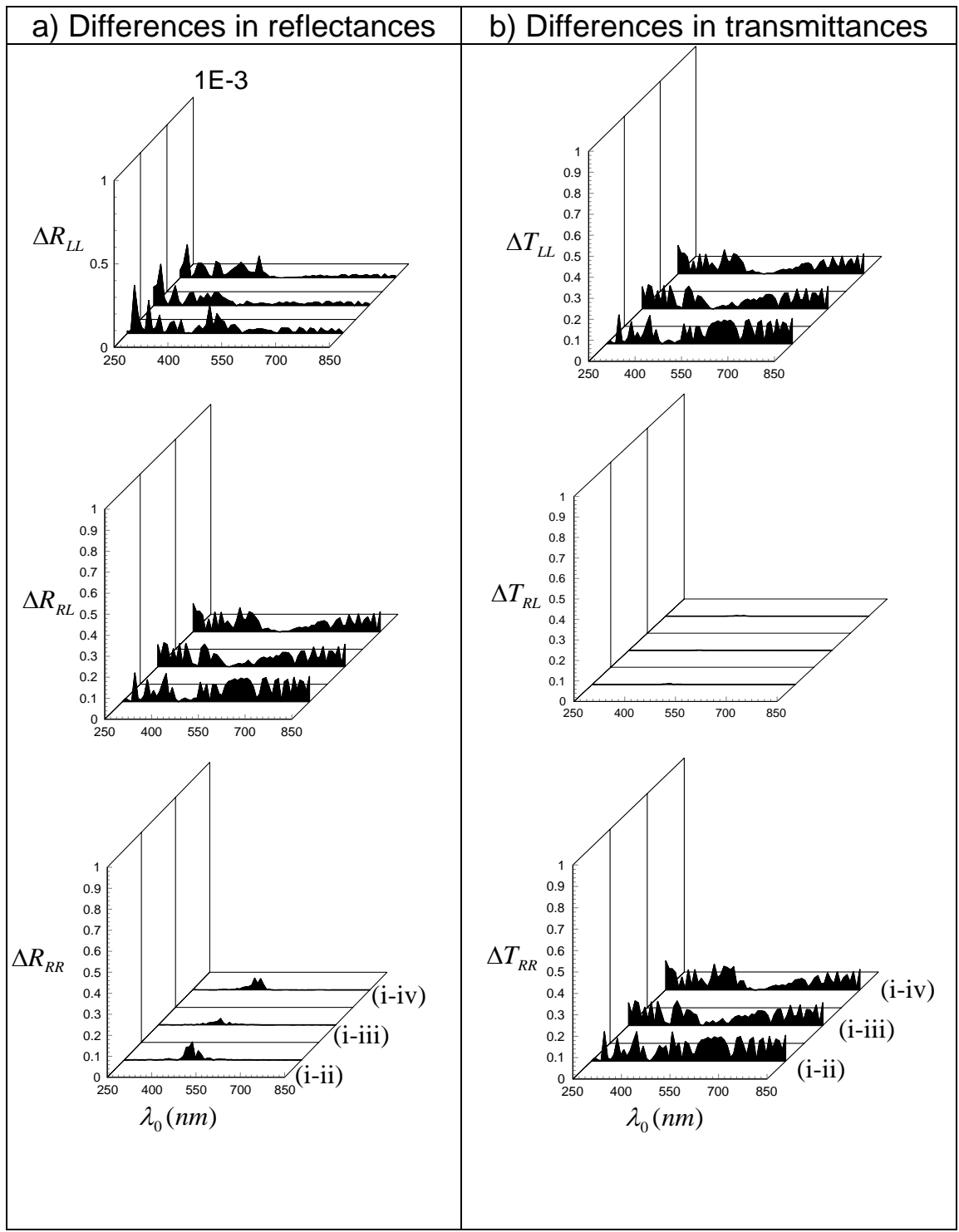

**Fig.3; F. Babaei and H. Savaloni**

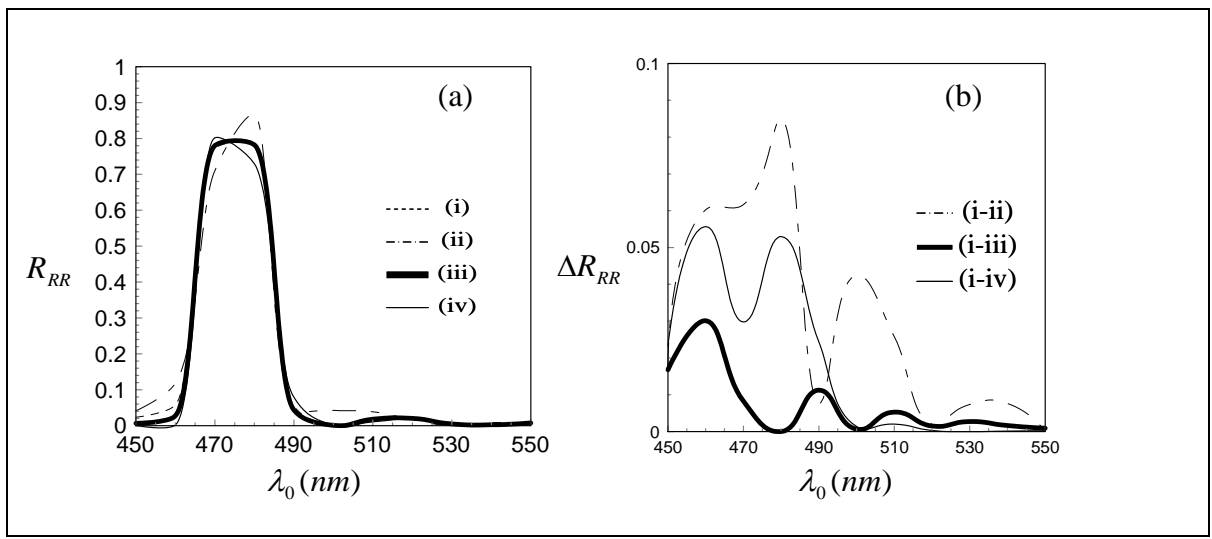

**Fig. 4.; F. Babaei and H. Savaloni**

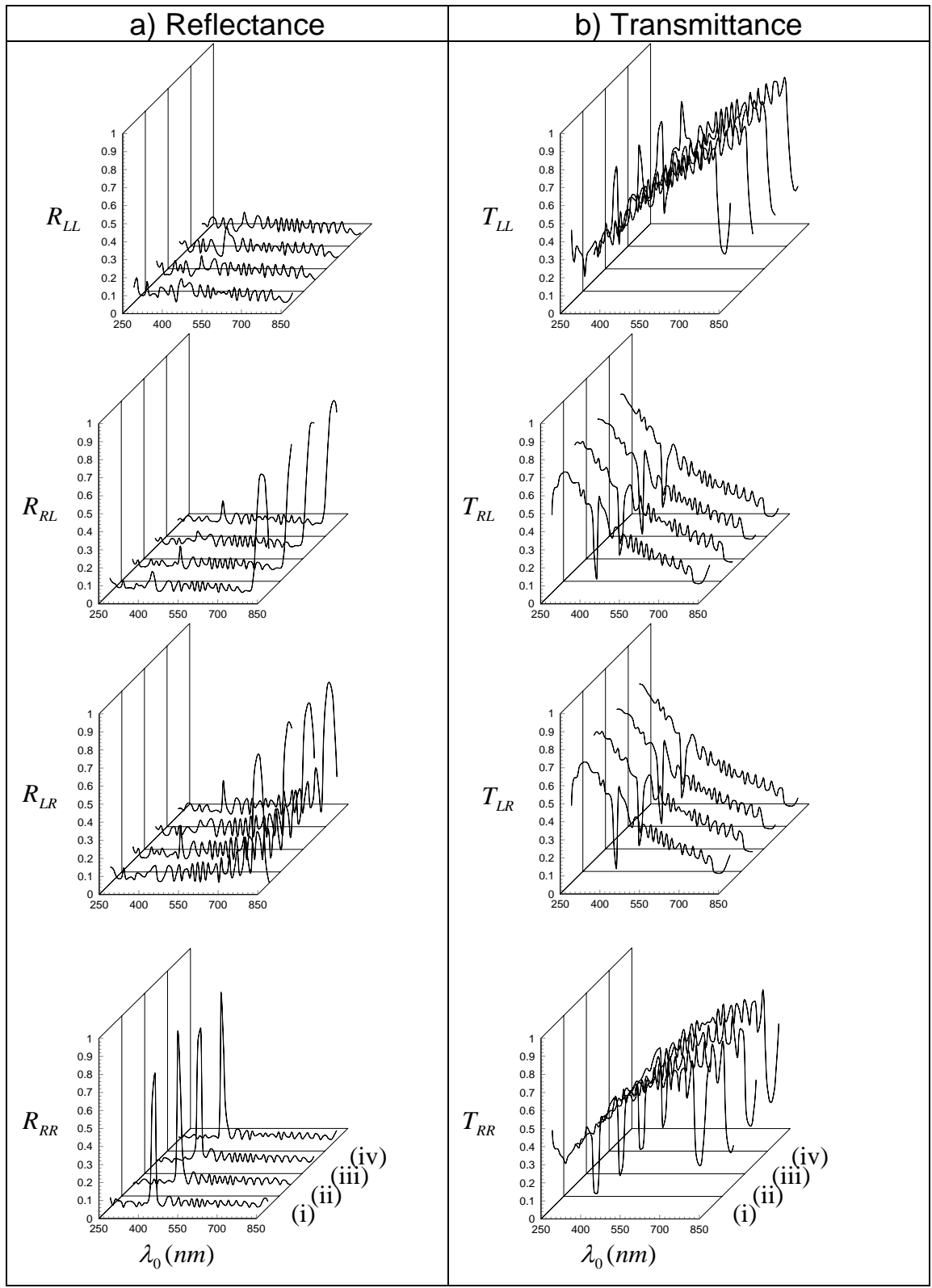

**Fig. 5; F. Babaei and H. Savaloni**

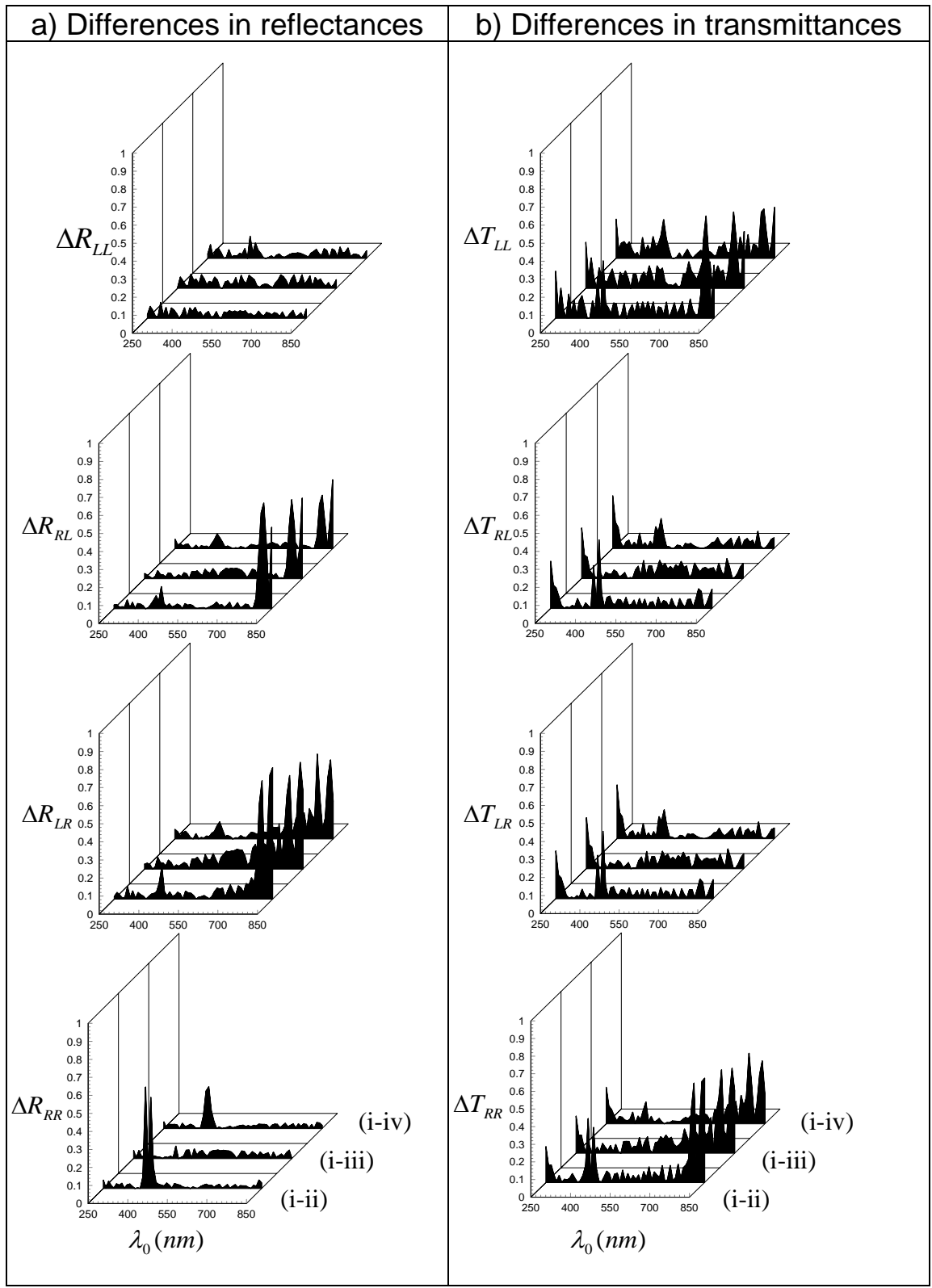

**Fig. 6; F. Babaei and H. Savaloni**

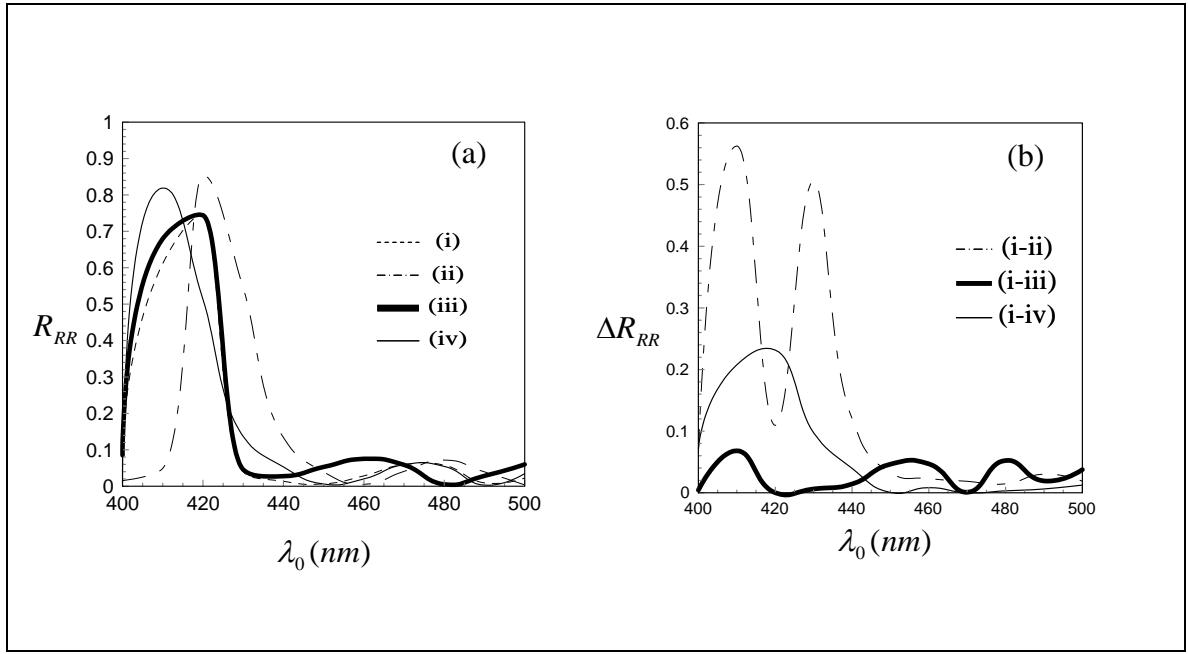

**Fig.7.; F. Babaei and H. Savaloni**

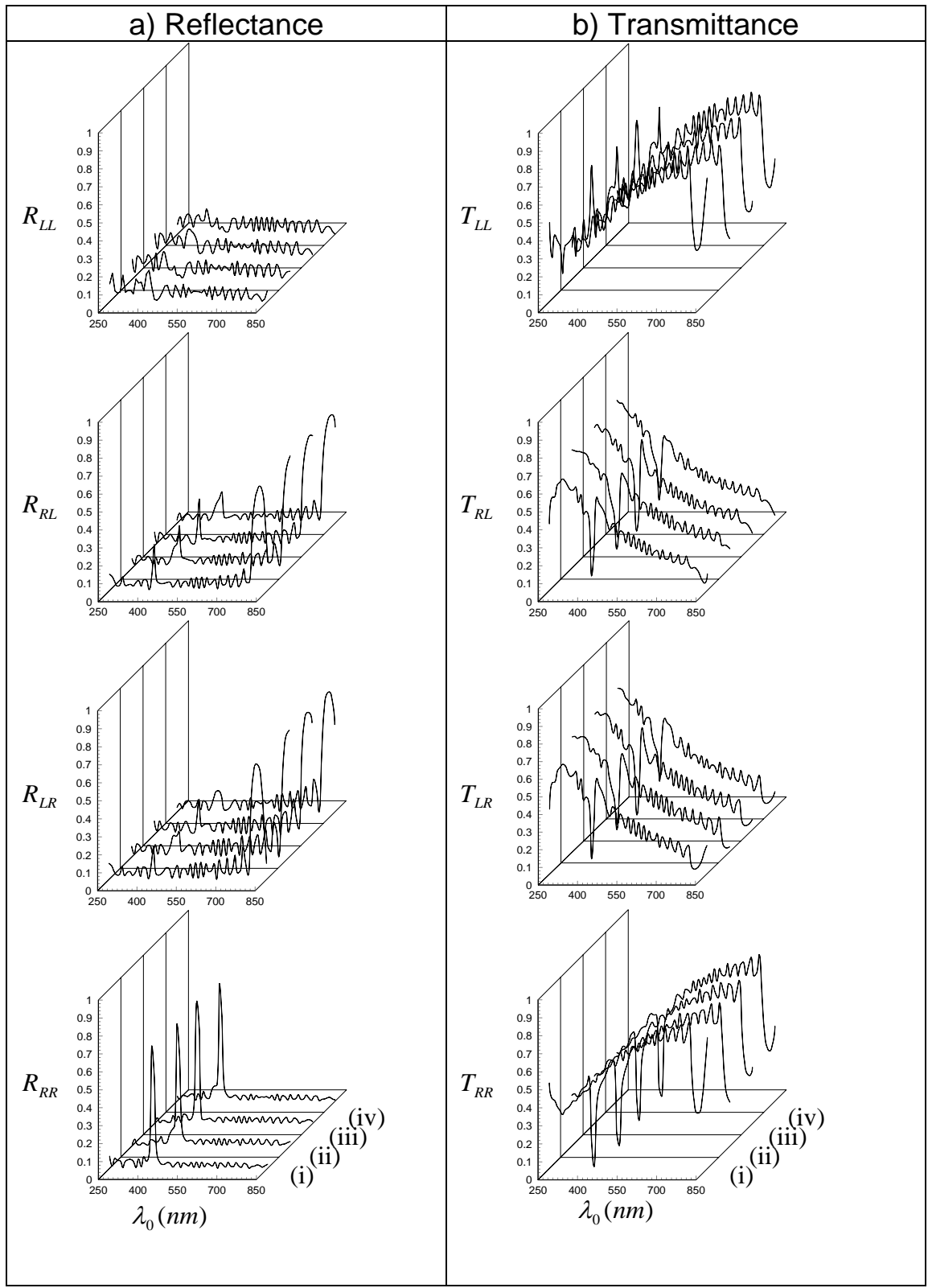

**Fig. 8; F. Babaei and H. Savaloni**

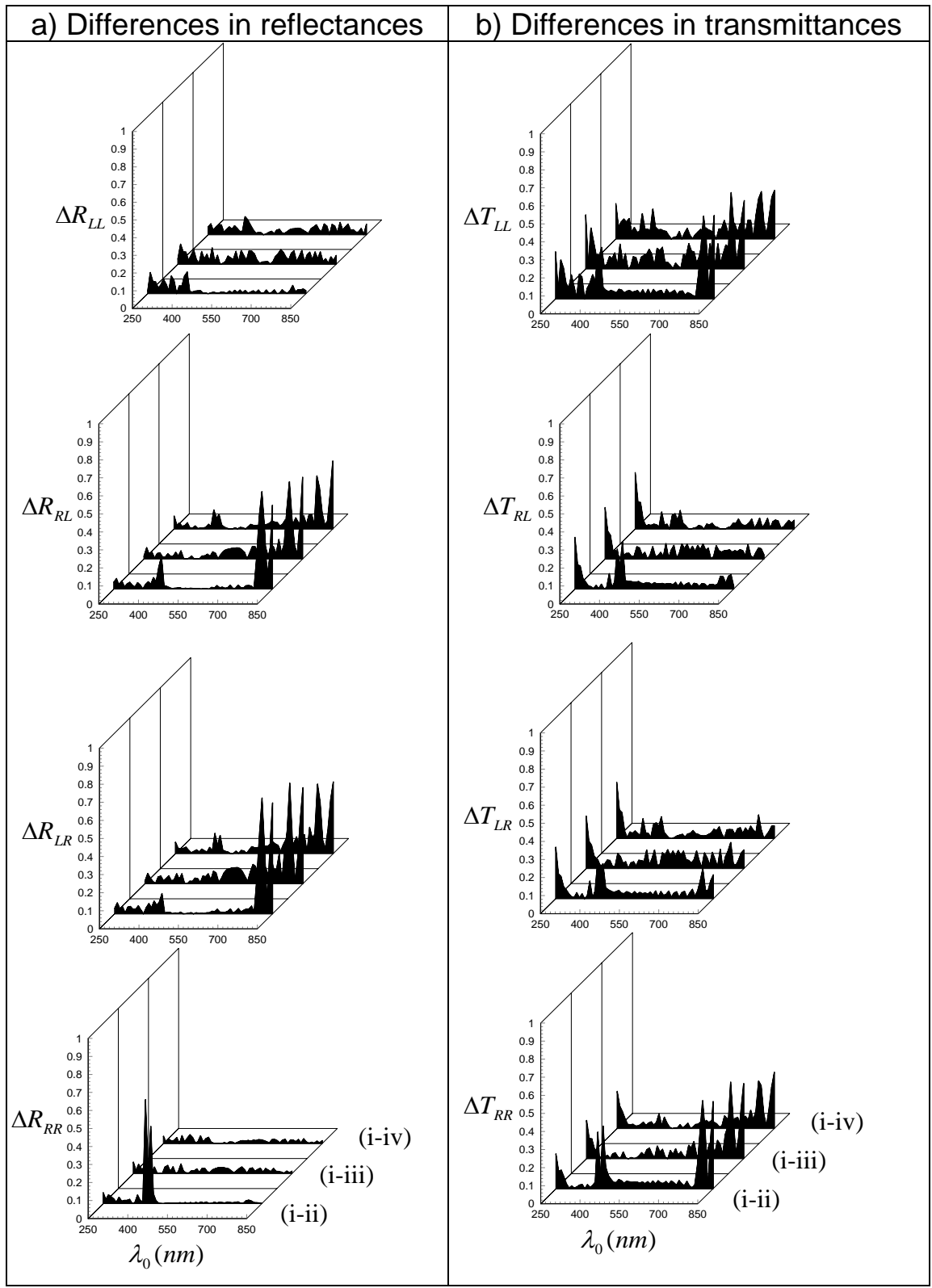

**Fig. 9; F. Babaei and H. Savaloni**

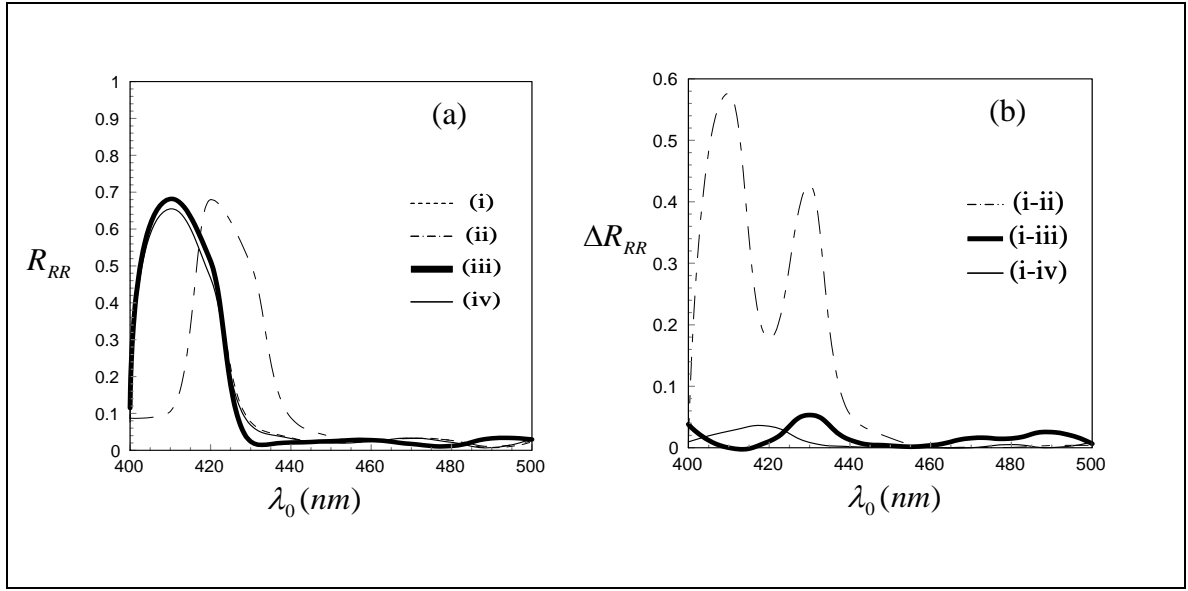

**Fig.10.; F. Babaei and H. Savaloni**